\documentclass[prl,aps,twocolumn,groupedaddress]{revtex4}

\usepackage{bm}
\usepackage{graphicx}
\usepackage{color}
\usepackage{ulem}

\begin{document}

\title{Multiple energy scales and anisotropic energy gap in the charge-density-wave phase of kagome superconductor CsV$_3$Sb$_5$}

\author{Kosuke Nakayama,$^{1,2,\#,\ast}$ Yongkai Li,$^{3,4,\#}$ Takemi Kato,$^{1}$ Min Liu,$^{3,4}$ Zhiwei Wang,$^{3,4,\ast}$ Takashi Takahashi,$^{1,5,6}$ Yugui Yao,$^{3,4}$ and Takafumi Sato$^{1,5,6,7}$}
\email[\#These authors equally contributed to this work.\\
$^{\ast}$Corresponding authors:]{k.nakayama@arpes.phys.tohoku.ac.jp; zhiweiwang@bit.edu.cn; t-sato@arpes.phys.tohoku.ac.jp}
\affiliation{$^1$Department of Physics, Faculty of Science, Tohoku University, Sendai 980-8578, Japan\\
$^2$Precursory Research for Embryonic Science and Technology (PRESTO), Japan Science and Technology Agency (JST), Tokyo, 102-0076, Japan\\
$^3$Centre for Quantum Physics, Key Laboratory of Advanced Optoelectronic Quantum Architecture and Measurement (MOE), School of Physics, Beijing Institute of Technology, Beijing 100081, China\\
$^4$Beijing Key Lab of Nanophotonics and Ultrafine Optoelectronic Systems, Beijing Institute of Technology, Beijing 100081, China\\
$^5$Center for Spintronics Research Network, Tohoku University, Sendai 980-8577, Japan\\
$^6$WPI Research Center, Advanced Institute for Materials Research, Tohoku University, Sendai 980-8577, Japan\\
$^7$International Center for Synchrotron Radiation Innov1ation Smart (SRIS), Tohoku University, Sendai 980-8577, Japan\\
}

\date{\today}

\begin{abstract}
Kagome metals $A$V$_3$Sb$_5$ ($A$ = K, Rb, and Cs) exhibit superconductivity at 0.9-2.5 K and charge-density wave (CDW) at 78-103 K. Key electronic states associated with the CDW and superconductivity remain elusive. Here, we investigate low-energy excitations of CsV$_3$Sb$_5$ by angle-resolved photoemission spectroscopy. We found an energy gap of 50-70 meV at the Dirac-crossing points of linearly dispersive bands, pointing to an importance of spin-orbit coupling. We also found a signature of strongly Fermi-surface and momentum-dependent CDW gap characterized by the larger energy gap of maximally 70 meV for a band forming a saddle point around the M point, the smaller (0-18 meV) gap for a band forming massive Dirac cones, and a zero gap at the $\Gamma$/A-centered electron pocket. The observed highly anisotropic CDW gap which is enhanced around the M point signifies an importance of scattering channel connecting the saddle points, laying foundation for understanding the nature of CDW and superconductivity in $A$V$_3$Sb$_5$.
\end{abstract}


\maketitle
Kagome lattice, consisting of $3d$ transition metal ions with two-dimensional (2D) network of corner-sharing triangles, provides an excellent platform to explore novel quantum phenomena originating from electron correlation and nontrivial band topology. While insulating kagome lattice has been intensively studied in relation to geometric frustration and quantum magnetism \cite{RamirezARMS1994, ShoresJACS2005, PollmannPRL2008, BalentsNature2010, YanScience2011, HanNature2012, HanPRL2012, FuScience2015}, metallic kagome lattice is currently attracting particular attention because of its unique band structure characterized by the nearly flat band and Dirac-cone band that promote strong-correlation and topological effects. Depending on the electron filling of the kagome-lattice bands, various intriguing quantum states have been predicted, e.g. Weyl magnet \cite{NayakSA2016, YangNJP2017, KurodaNM2017, YeNature2018, LiuNP2018, MoraliScience2019, LiuScience2019}, density wave orders \cite{IsakovPRL2006, GuoPRB2009, WangPRB2013}, charge fractionalization \cite{OBrienPRB2010, RueggPRB2011}, and superconductivity \cite{WangPRB2013, KoPRB2009, KieselPRB2012, KieselPRL2013}.

Recently, a family of $A$V$_3$Sb$_5$ ($A$ = K, Rb, and Cs) was discovered to be a kagome superconductor \cite{OrtizPRM2019, OrtizPRL2020, OrtizPRM2021} with superconducting transition temperature $T_{\rm c}$ of 0.93-2.5 K, despite the fact that kagome metals rarely become a superconductor. $A$V$_3$Sb$_5$ crystallizes in a layered structure consisting of alternately stacked V kagome-lattice layer with hexagonally arranged Sb atoms (V1 and Sb1), graphene-like Sb layer (Sb2), and hexagonal $A$ layer [see Fig. 1(a)]. Besides superconductivity, $A$V$_3$Sb$_5$ commonly undergoes a charge-density wave (CDW) transition at $T_{\rm CDW}$ = 78-103 K accompanied with three-dimensional (3D) 2$\times$2$\times$2 charge order \cite{LiAX2021, LiangAX2021}. Angle-resolved photoemission spectroscopy (ARPES) clarified that the kagome-lattice bands participate in the states near the Fermi level ($E_{\rm F}$), by observing the Dirac-cone-like bands forming a large hexagonal Fermi surface (FS) centered at the $\Gamma$ point and a saddle point near $E_{\rm F}$ at the M point, together with an electron pocket at the $\Gamma$ point of 2D Brillouin zone (BZ), consistent with the density-functional-theory (DFT) calculations \cite{OrtizPRL2020, YangSA2020, LiAX2021, LiuAX2021}.

While overall electronic structure of $A$V$_3$Sb$_5$ is almost established, the mechanism of superconductivity and CDW is highly controversial. $Z_2$ topological invariant in the normal state suggested by the DFT calculation \cite{OrtizPRL2020, OrtizPRM2021} may point to unconventional (topological) superconductivity. Existence of a saddle point in the calculated band structure was discussed to promote the $d$-wave pairing associated with the scattering between saddle points via $Q$ = ($\pi$, 0) vector \cite{NandkishoreNP2012}. Weak electron-phonon coupling suggested by the DFT calculation \cite{TanAX2021} also supports an unconventional pairing. In the experiment, conventional vs. unconventional nature of superconductivity is far from reaching a consensus, as represented by contradictory reports on the presence/absence of gap nodes \cite{DuanAX2021, CZhaoAX2021}. And very recently, a nematic electronic state and a two-fold symmetry superconductivity was observed in CsV$_3$Sb$_5$ \cite{Xiang AX2021}.

The electronic states relevant to the CDW formation are also under intensive debate. Apparent CDW-gap opening was not observed in previous ARPES studies on RbV$_3$Sb$_5$ and CsV$_3$Sb$_5$ \cite{OrtizPRL2020, LiuAX2021}, whereas scanning tunneling microscopy/spectroscopy (STM/STS) on KV$_3$Sb$_5$ and CsV$_3$Sb$_5$ reported a gap opening \cite{JiangAX2020, ChenAX2021, HZhaoAX2021}. While the calculation suggested that the CDW is triggered by the Peierls instability due to the FS nesting \cite{TanAX2021}, x-ray scattering on RbV$_3$Sb$_5$ and CsV$_3$Sb$_5$ suggested an electronic-driven mechanism because of the absence of expected phonon anomaly \cite{LiAX2021}. In contrast, optical spectroscopy on CsV$_3$Sb$_5$ suggested the FS-nesting mechanism by observing the reduction of Drude weight below $T_{\rm CDW}$ for the saddle-point-related feature \cite{ZhouAX2021}. Thus, the mechanism of CDW and superconductivity is still far from being established.

\begin{figure}
\includegraphics[width=3.4in]{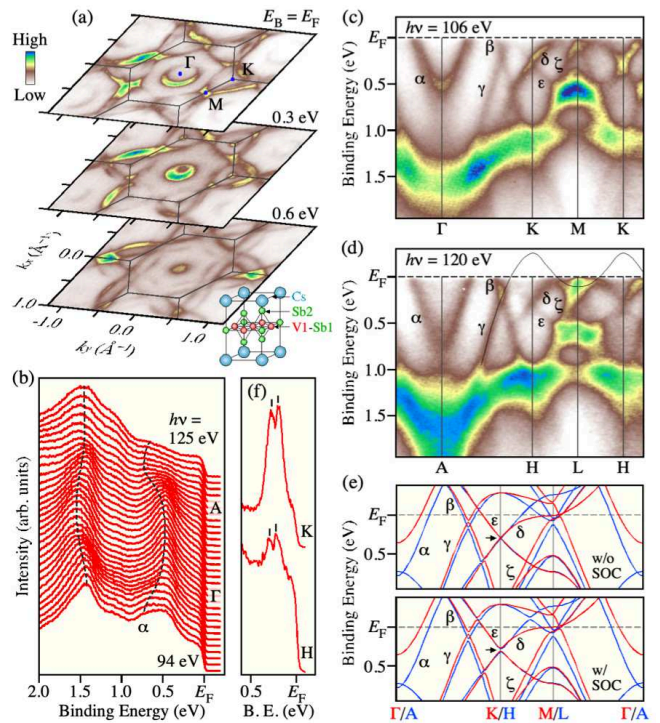}
\vspace{0cm}
\caption{(a) ARPES-intensity maps at binding energies ($E_{\rm B}$'s) of $E_{\rm F}$, 0.3 eV, and 0.6 eV plotted as a function of $k_x$ and $k_y$, for CsV$_3$Sb$_5$ measured at $T$ = 20 K with 106-eV photons. (b) Photon-energy dependence of the normal-emission energy distribution curves (EDCs). The inner potential $V_0$ was estimated to be 10 eV. (c), (d) ARPES intensity as a function of wave vector and $E_{\rm B}$ measured along the $\Gamma$KM ($k_z$ = 0) and AHL ($k_z$ = $\pi$) cuts, respectively. Solid curve in (d) is a guide to the eyes to trace the experimental $\gamma$ band dispersion. (e) Calculated band structure along the $\Gamma$KM$\Gamma$ (AHLA) cut (top) without and (bottom) with SOC. Calculated $E_{\rm F}$ is shifted downward by 50 meV to obtain a better matching with the experimental data. (f) EDCs at the Dirac-crossing points.}
\end{figure}

In this article, we report a high-resolution ARPES study of CsV$_3$Sb$_5$ single crystals. We established the low-energy excitations in the CDW phase, and found (i) the gap opening at the Dirac points due to the spin-orbit coupling (SOC) and (ii) the strongly FS and momentum ($k$) dependent CDW gap characterized by multiple energy scales. We discuss implications of the present results in relation to the mechanism of CDW and superconductivity in $A$V$_3$Sb$_5$.

We at first present the overall band structure of CsV$_3$Sb$_5$ (see Supplemental Material for details of the experimental condition and the band-calculation method \cite{SM}). Figure 1(a) shows the ARPES-intensity map at representative binding energy ($E_{\rm B}$) slices as a function of $k_x$ and $k_y$, obtained with 106-eV photons which probe the $k_z$ $\sim$ 0 plane of the bulk BZ [the $k_z$ value was determined by the periodicity of band dispersion as a function of photon energy ($h\nu$); $h\nu$ variation of the band in the normal-emission measurement is highlighted in Fig. 1(b) and Supplementary Fig. 5 \cite{SM}]. We found that the overall intensity pattern in Fig. 1(a) is consistent with the previous ARPES reports on $A$V$_3$Sb$_5$ \cite{OrtizPRL2020, YangSA2020, LiAX2021, LiuAX2021}. One can identify a circular pocket centered at the $\Gamma$ point. This pocket originates from an electron band (called here the $\alpha$ band) as visible from ARPES-intensity plot along the $\Gamma$KM cut in Fig. 1(c). According to the DFT calculations \cite{LiuAX2021, JiangAX2020, JZhaoAX2021}, this band is attributed to the $5p_z$ band of Sb1 atoms embedded in the kagome-lattice plane. One can recognize in Fig. 1(a) triangular shaped intensity pattern centered at each K point which connects to each other around the M point and forms a large hexagonal FS centered at the $\Gamma$ point. As shown in Fig. 1(c), this pocket originates from a band located at 1 eV at the K point (the $\beta$ band) that displays an overall linear dispersion toward $E_{\rm F}$ on approaching the $\Gamma$ point and crosses $E_{\rm F}$ at $k \sim 2/3$ of the $\Gamma$K interval. This band is reproduced in the calculation in Fig. 1(e) and attributed to the kagome-lattice band with mainly V-$3d_{xy}$ character \cite{LiuAX2021, JZhaoAX2021}. This band intersects other two linearly dispersive bands ($\gamma$ and $\delta$ bands) at $\sim$0.1 eV and $\sim$0.5 eV, and the $\delta$ band also intersects the linearly dispersive $\epsilon$ band at the K point, forming multiple Dirac points (note that the Dirac point associated with the $\beta$-$\gamma$ band crossing is predicted to form a nodal line along $k_z$ when the SOC is neglected \cite{JZhaoAX2021}). The $\delta$ band rapidly disperses toward $E_{\rm F}$ on approaching the M point and participates in forming the saddle point. At the M point, there is another holelike band ($\zeta$ band) with the top of dispersion below $E_{\rm F}$.

At $h\nu$ = 120 eV that corresponds to the $k_z$ = $\pi$ plane, there are several similarities in the band dispersions with those at the $k_z$ = 0 plane, e.g. linear dispersions of the $\beta$, $\gamma$, $\delta$, and $\epsilon$ bands and their Dirac-cone formation [compare Figs. 1(c) and 1(d)]. Besides such similarities, one can recognize a clear difference in energy position of the $\alpha$-band bottom, i.e. it is located at $E_{\rm B}$ $\sim$ 0.5 and 0.75 eV at $k_z$ = 0 and $\pi$, respectively. One can also find that a shallow electron pocket centered at the L point appears only at the $k_z$ = $\pi$ plane. According to the calculation [Fig. 1(e)], this band is connected to the $\gamma$ band as indicated by a guide in Fig. 1(d). The observed band structures including finite $k_z$ dispersions reasonably agree with the bulk band calculation, supporting their bulk nature. It is noted that, while the existence of topological surface state at the $\bar{\rm M}$ point was predicted \cite{OrtizPRL2020, OrtizPRM2021}, we found no evidence for such a surface state below $E_{\rm F}$.

As shown in Fig. 1(c), the intensity of the $\delta$ band is weakened at which the $\epsilon$ band intersects at the K point at $\sim$0.3 eV. This is also the case for the band-crossing point at the H point [Fig. 1(d)]. To better visualize the band modulation associated with the intensity weakening, we plot in Fig. 1(f) the EDCs at the K and H points. These EDCs commonly exhibit two-peaked structure around the Dirac point, signifying an energy gap (called Dirac gap; see also Fig. 6 of Supplemental Material \cite{SM}). From the energy separation of two peaks, we have estimated the magnitude of Dirac gap to be 50-70 meV. The Dirac gap is associated with the SOC because the calculations predict a spin-orbit gap at the Dirac point (note that, by referring to the calculation without SOC, the gap at the K and H points is purely of SOC origin, whereas those at other \textbf{k} points are additionally affected by the band hybridization). Given the fact that some Dirac points are predicted to be located at very close to $E_{\rm F}$ in $A$V$_3$Sb$_5$ \cite{OrtizPRM2019}, our observation suggests that the SOC needs to be taken into account when the connection between the Dirac-fermion and transport properties (such as anomalous Hall effect \cite{YangSA2020, YuAX2021}) is discussed.

\begin{figure}
\includegraphics[width=3.4in]{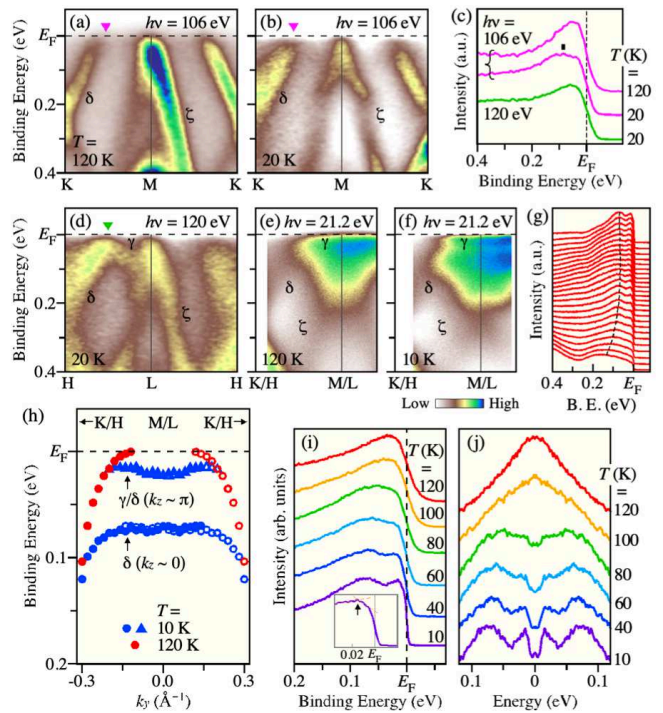}
\vspace{0cm}
\caption{(a), (b) ARPES intensity along the MK cut ($h\nu$ = 106 eV) measured at $T$ = 120 K and 20 K, respectively. (c) EDCs at the \textbf{k}$_{\rm F}$ point  of the $\delta$ band along the MK cut (magenta) and the $\gamma$ band along the LH cut (green). (d) Same as (b) but measured along the LH cut ($h\nu$ = 120 eV). (e), (f) Same as (a) and (b) but measured at 21.2-eV photons. (g) Corresponding EDCs of (f). (h) Experimental band dispersion at $T$ = 10 and 120 K extracted by tracing the peak position of EDCs in (e) and (f). Filled and open circles represent original and symmetrized data. (i) Temperature dependence of the EDC at the \textbf{k}$_{\rm F}$ point of shallow electron band measured across $T_{\rm CDW}$. Inset shows the magnified view for the EDC at 10 K. Black arrow indicates the peak position. (j) Same as (i) but symmetrized with respect to $E_{\rm F}$.}
\end{figure}

Next we present the electronic states associated with the CDW. Figures 2(a) and 2(b) show the ARPES-intensity plot near $E_{\rm F}$ measured along the MK cut at temperature above (120 K) and below (20 K) $T_{\rm CDW}$, respectively. In the normal state [Fig. 2(a)], the $\delta$ band crosses $E_{\rm F}$ at the midway of the MK line and forms a saddle point slightly above $E_{\rm F}$ at the M point. There also exists the $\zeta$ band which forms another saddle point at the M point, while it does not cross $E_{\rm F}$ along the MK cut. In the CDW phase [Fig. 2(b)], the intensity of the $\zeta$ band is modulated at $E_{\rm B}$ $\sim$ 0.2 eV likely due to band hybridization by CDW. More importantly, a hump structure at $E_{\rm B}$ $\sim$ 70 meV is observed at the \textbf{k}$_{\rm F}$ point of the $\delta$ band [see magenta curves in Fig. 2(c); see also Fig. 7 of Supplementary Material \cite{SM}]. Since the \textbf{k}$_{\rm F}$ point of the $\delta$ band is well isolated from the $\zeta$-band top, the 70-meV hump is unlikely attributed to the saddle point of the $\zeta$ band but most probably reflects a CDW-gap  opening on the $\delta$ band. To clarify the 3D nature of CDW, the near-$E_{\rm F}$ ARPES intensity at 20 K along the LH cut is plotted in Fig. 2(d). Here, a shallow $\gamma$ electron pocket is seen around $E_{\rm F}$ near the \textbf{k}$_{\rm F}$ point of the $\delta$ band. Intriguingly, unlike the MK cut, EDC near \textbf{k}$_{\rm F}$ of the $\gamma$/$\delta$ band does not show a clear hump structure [green curve in Fig. 2(c)], suggesting the bulk origin of the hump structure and the reduction of the CDW-gap size around the LH line.

To gain further insights into the electronic structure in the CDW phase, we have performed higher-resolution ($\Delta E$ = 7 meV) measurements by utilizing low-energy photons ($h\nu$ = 21.2 eV) [Figs. 2(e)-2(f)]. We found that, while 21.2-eV photons are expected to probe $k_z$ $\sim$ 0.8$\pi$ by referring to the estimated $V_0$ value, the ARPES data simultaneously reflect the band structures at around $k_z$ = 0 and $\pi$ due to a $k_z$ broadening effect (for details, see Fig. 8 of Supplemental Material \cite{SM}). In the CDW phase, one can see a broad hump in the EDCs at $\sim$150 meV around the K/H point [dashed curve in Fig. 2(g)] which disperses toward $E_{\rm F}$ on approaching the M/L point and stays at $\sim$70 meV around the M/L point to form a nearly flat band. This hump feature is attributed to the gapped $\delta$ band at $k_z$ = 0 seen in Figs. 2(b) and 2(c). There exists another sharp peak near $E_{\rm F}$, separated from the hump by a dip at 35 meV. This band is assigned to the $\gamma$ electron pocket at $k_z$ = $\pi$ with a finite mixture of the $\delta$ band ($k_z$ = $\pi$ component) near \textbf{k}$_{\rm F}$. In fact, as shown by a plot of experimental band dispersion at 10 K obtained by tracing the peak position of EDCs in Fig. 2(h), the low-energy branch has an electronlike dispersion. Thus, the characteristic peak-dip-hump structure in the EDCs is attributed to the existence of two types of bands at different $k_z$'s, the hump and peak, called here the saddle-point band and Dirac band, respectively. We found that the peak-dip-hump structure vanishes above $T_{\rm CDW}$ [Fig. 2(e); for the experimental band dispersion, see red circles in Fig. 2(h)]. To examine whether this change is associated with the CDW, we show in Fig. 2(i) the temperature dependence of EDC at the \textbf{k}$_{\rm F}$ point of Dirac band. The hump survives up to 80 K (below $T_{\rm CDW}$) and disappears at 100 K (above $T_{\rm CDW}$), demonstrating that the hump is indeed associated with the formation of CDW and reflects the CDW gap for the saddle-point band.

An important finding manifests itself by a careful inspection of the EDCs in Fig. 2(i). The sharp peak at $T$ = 10 K is not located at $E_{\rm F}$ but at slightly higher $E_{\rm B}$ ($\sim$15 meV) as seen from the magnified EDCs in the inset (see arrow). Symmetrized EDC in Fig. 2(j) reveals spectral-weight suppression in the narrow energy region of $\pm$15 meV within the sharp peaks, indicative of the small gap opening at the Dirac band (note that this small gap is not resolved in Figs. 2(b) and 2(c) because of insufficient energy resolution for higher $h\nu$'s). This gap is associated with the CDW, because it is gradually weakened on elevating temperature and disappears above $T_{\rm CDW}$, in accordance with the behavior of saddle-point band at 70 meV. These results indicate that the CDW of CsV$_3$Sb$_5$ is characterized by at least two energy scales.

\begin{figure}
\includegraphics[width=3.4in]{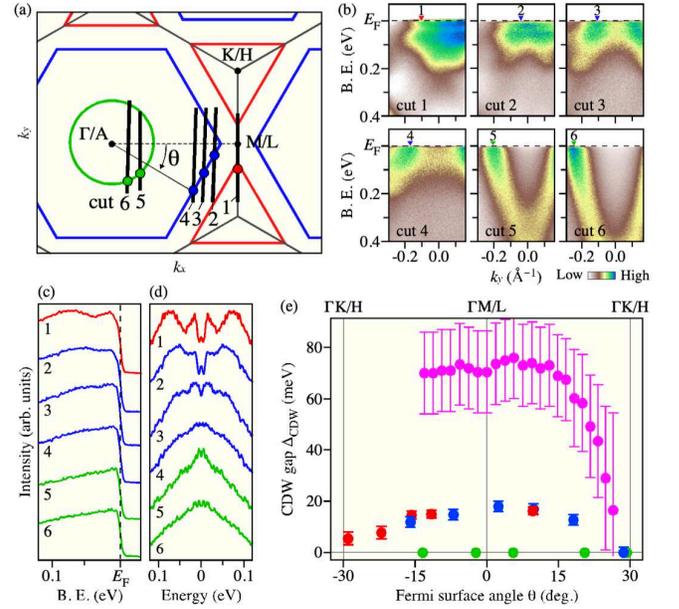}
\vspace{0cm}
\caption{(a) Schematic FS in the surface BZ together with \textbf{k} cuts (cuts 1-6) and \textbf{k} points (points 1-6) where the intensity in (b) and the EDCs in (c) were obtained. Definition of the FS angle $\theta$ is also indicated. (b) ARPES intensity measured at $T$ = 10 K along cuts 1-6 in (a). (c), (d) EDCs and symmetrized EDCs at $T$ = 10 K, respectively, at points 1-6 in (a). (e) Plots of the CDW-gap magnitude $\Delta_{\rm CDW}$ as a function of $\theta$. Magenta, red, blue, and green circles represent the CDW gaps for the saddle-point band, triangular FS centered at the K/H point, hexagonal FS centered at the $\Gamma$/A point, and electron band at the $\Gamma$/A, respectively.}
\end{figure}

Throughout the whole 2D BZ, we have investigated the CDW gap for the saddle-point and Dirac bands, and found that the gap is strongly \textbf{k} dependent. Figure 3(b) shows the ARPES intensity plotted along several \textbf{k} cuts shown as cuts 1-6 in the FS mapping in Fig. 3(a) which covers the \textbf{k}$_{\rm F}$ points of the shallow electron band (cut 1), the hexagonal pocket associated with the Dirac band (cuts 2-4), and the electron pocket at $\Gamma$/A (cuts 5-6). The dip-originated spectral weight suppression, which is a measure of the CDW gap opening for the saddle-point band, is most prominent along cut 1 (MK/LH cut) and gradually becomes less prominent on moving away from the M/L point, as visible from a systematic change in the ARPES intensity along cuts 2-4. We have extracted the EDCs at representative \textbf{k}$_{\rm F}$ points on the Dirac band (points 1-4) in Fig. 3(c), and found that the hump gradually approaches $E_{\rm F}$ on moving away from the M point. This signifies the strong \textbf{k} dependence of the CDW gap. It is noted that the temperature-induced band energy shift in Fig. 2(h) is another measure of the gap size for the saddle-point band; one can confirm that the band shift becomes smaller (i.e., the gap becomes smaller) as moving away from the M point. The symmetrized EDC at $T$ = 10 K in Fig. 3(d) further reveals that the Dirac band near $E_{\rm F}$ also shows anisotropic CDW gap; the gap is $\sim$16 meV at point 1 but it vanishes at point 4 (along the $\Gamma$K/AH line). Another important finding is that there exists no obvious spectral anomaly for the $\Gamma$/A-centered electron pocket [see cuts 5 and 6 in Fig. 3(b)]. Also, the hump is absent at points 5 and 6 in Figs. 3(c) and 3(d). These results indicate that the CDW gap is absent in the electron pocket at $\Gamma$/A. We have estimated the size of CDW gap $\Delta_{\rm CDW}$ at $T$ = 10 K at various \textbf{k}$_{\rm F}$ points and plotted the $\Delta_{\rm CDW}$ as a function of FS angle $\theta$ in Fig. 3(e) (see Fig. 9 of Supplemental Material for a full data set \cite{SM}). As clearly seen, the CDW gap is larger for the saddle-point band (maximally 70 meV) and smaller for the Dirac band (maximally 18 meV), whereas it is absent on entire electron pocket at $\Gamma$/A. Intriguingly, the CDW gap has a strong anisotropy and takes a maximum around $\theta$ = 0$^{\circ}$ (along the $\Gamma$M/AL cut) and a minimum around $\theta$ = $\pm$30$^{\circ}$ (along the $\Gamma$K/AH cut) for both the saddle-point and Dirac bands.

The ARPES-determined CDW gap has a good correspondence to the unusual behavior in the DOS observed by other spectroscopic techniques. STM/STS reported a broad hump feature at $\sim$70 meV and V-shaped DOS within $\pm$10-20 meV of $E_{\rm F}$ at low temperatures \cite{JiangAX2020, ChenAX2021, HZhaoAX2021}. These features likely correspond to the CDW gaps for the saddle-point and Dirac bands, respectively. STM/STS also found residual DOS at zero bias voltage, and it could be contributed from the electron pocket at $\Gamma$ and the gapless portion of the Dirac band. Recent optical spectroscopy of CsV$_3$Sb$_5$ revealed a marked suppression of the Drude weight below $T_{\rm CDW}$ \cite{ZhouAX2021}. This would be also related to the CDW gap for the saddle-point band because its energy scale ($\Delta_{\rm CDW}$ $\sim$86.5 meV) is similar to the hump energy in ARPES. Thus, our ARPES data definitely help interpreting characteristic spectroscopic signatures seen by other experiments.

Our observation also sheds light on the mechanism of CDW and superconductivity in $A$V$_3$Sb$_5$. Theoretical studies predicted that the inter-band scattering between saddle points promote CDW or unconventional density waves \cite{TanAX2021, FengAX2021}, but direct experimental proof was elusive. The observed large CDW gap at the saddle-point band around the $\bar{\rm M}$ point supports this prediction and suggests that the electron scattering via $Q$ = ($\pi$, 0) connecting saddle points contributes largely to the energy gain for stabilizing the CDW with in-plane 2$\times$2 component. To understand this point in more detail, we plot a schematic of the reconstructed FS in Fig.4. Original and folded FSs are indicated by solid and dashed curves, respectively, with orbital-dependent colorings [original FSs were determined by tracing the ARPES intensity at $E_{\rm F}$ in Fig. 4(a)]. Also, original and folded saddle-point regions of the $\delta$ band are shown by magenta and light magenta paintings, respectively. One can see that original and folded saddle points around the M point overlap well, whereas the FSs along the $\Gamma$K cut show a finite mismatch as marked by red arrows. Such a difference in the band/FS overlaps via the 2$\times$2 $Q$ vector qualitatively explains the observed CDW-gap anisotropy; the gap size is larger around the M point due to the overlap of the original and folded saddle points. A remaining issue is the origin of additional 3D component of CDW (with 2$\times$2$\times$2 periodicity), which is puzzling because the CDW vector connects the M point to the L point where no saddle point exists. This necessitates further $k_z$ selective experiments. It is also predicted that the scattering between saddle points may promote unconventional superconductivity, depending on the degree of competition with CDW \cite{NandkishoreNP2012}. However, the observed large CDW gap suggests substantial reduction of DOS responsible for the superconducting pairing on the saddle-point band. Microscopic theories of superconductivity must be constructed based on such a largely gapped saddle-point band and the CDW-gap nodes for the Dirac bands. In this regard, one can even suggest that major superconducting carriers could originate from the Sb $5p_z$ bands at the $\bar{\Gamma}$ point rather than the V $3d$ bands participating in the 2D kagome network.

\begin{figure}
\includegraphics[width=3.4in]{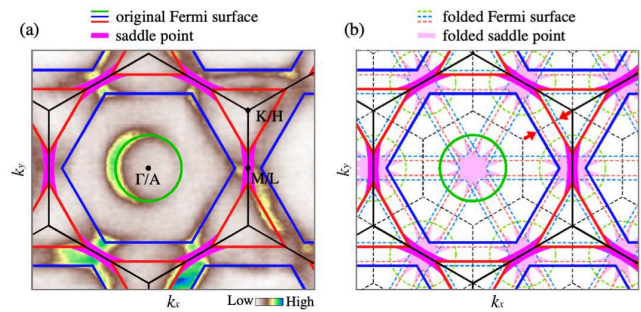}
\vspace{0cm}
\caption{(a) ARPES intensity at $E_{\rm F}$ measured at the $k_z$ $\sim$ 0 plane, together with the experimental FSs (solid curves) and the momentum region where the saddle-point of the $\delta$ band is present (magenta). (b) Original FSs extracted in (a) (solid curves) and their folded counterparts by the 2$\times$2 CDW wave vector (dashed curves). Original and folded saddle-point regions are shown by magenta and light magenta, respectively.}
\end{figure}

In conclusion, the present ARPES study of CsV$_3$Sb$_5$ has revealed two key signatures in the band structure, (i) the energy gap on the multiple Dirac points due to the SOC, and (ii) the modification of band structure associated with the CDW-gap opening. We found that the CDW-induced band modification is categorized into three types, the larger CDW gap for the saddle point around the M point, the smaller CDW gap for the Dirac band, and the absence of CDW gap for the electron band at $\Gamma$/A. The CDW gap also displays a strong anisotropy. The present result opens a pathway toward understanding the mechanisms of CDW and superconductivity in $A$V$_3$Sb$_5$.

Note Added: During the preparation of this manuscript, we became aware of \cite{WangAX2021}, which also reports a CDW gap opening on CsV$_3$Sb$_5$.

\begin{acknowledgments}
This work was supported by JST-CREST (No. JPMJCR18T1), JST-PRESTO (No. JPMJPR18L7), and Grant-in-Aid for Scientific Research (JSPS KAKENHI Grant No. JP17H01139, JP18H01160). The work at Beijing was supported by the National Key R\&D Program of China (Grant No. 2020YFA0308800), the Natural Science Foundation of China (Grant No. 92065109), the Beijing Natural Science Foundation (Grant No. Z190006), and the Beijing Institute of Technology (BIT) Research Fund Program for Young Scholars (Grant No. 3180012222011). Z.W thanks the Analysis \& Testing Center at BIT for assistances in facility support.
\end{acknowledgments}

\bibliographystyle{prsty}

\begin{thebibliography}{50}
\bibitem{RamirezARMS1994} A. P. Ramirez, Annu. Rev. Mater. Sci. \textbf{24}, 453-480 (1994).
\bibitem{ShoresJACS2005} M. P Shores, E. A Nytko, B. M. Bartlett, and D. G. Nocera, J. Am. Chem. Soc. \textbf{127}, 13462-13463 (2005).
\bibitem{PollmannPRL2008} F. Pollmann, P. Fulde, and K. Shtengel, Phys. Rev. Lett. \textbf{100}, 136404 (2008).
\bibitem{BalentsNature2010} L. Balents, Nature \textbf{464}, 199-208 (2010).
\bibitem{YanScience2011} S. Yan, D. A. Huse, and S. R. White, Science \textbf{332}, 1173-1176 (2011).
\bibitem{HanNature2012} T.-H. Han, J. S. Helton, S. Chu, D. G. Nocera, J. A. Rodriguez-Rivera, C. Broholm, and Y. S. Lee, Nature \textbf{492}, 406-410 (2012).
\bibitem{HanPRL2012} T. Han, S. Chu, and Y. S. Lee, Phys. Rev. Lett. \textbf{108}, 157202 (2012).
\bibitem{FuScience2015} M. Fu, T. Imai, T.-H. Han, and Y. S. Lee, Science \textbf{350}, 655-658 (2015).
\bibitem{NayakSA2016} A. K. Nayak, J. E. Fischer, Y. Sun, B. Yan, J. Karel, A. C. Komarek, C. Shekhar, N. Kumar, W. Schnelle, J. K. bler, C. Felser, and S. S. P. Parkin, Sci. Adv. \textbf{2}, e1501870 (2016).
\bibitem{YangNJP2017} H. Yang, Y. Sun, Y. Zhang, W.-J. Shi, S. S. P. Parkin, and B. Yan, New J. Phys. \textbf{19}, 015008 (2017).
\bibitem{KurodaNM2017} K. Kuroda, T. Tomita, M.-T. Suzuki, C. Bareille, A. A. Nugroho, P. Goswami, M. Ochi, M. Ikhlas, M. Nakayama, S. Akebi, R. Noguchi, R. Ishii, N. Inami, K. Ono, H. Kumigashira, A. Varykhalov, T. Muro, T. Koretsune, R. Arita, S. Shin, T. Kondo, and S. Nakatsuji, Nat. Mater. \textbf{16}, 1090-1095 (2017).
\bibitem{YeNature2018} L. Ye, M. Kang, J. Liu, F. v. Cube, C. R. Wicker, T. Suzuki, C. Jozwiak, A. Bostwick, E. Rotenberg, D. C. Bell, L. Fu, R. Comin, and J. G. Checkelsky, Nature \textbf{555}, 638-642 (2018).
\bibitem{LiuNP2018}E. Liu, Y. Sun, N. Kumar, L. Muechler, A. Sun, L. Jiao, S.-Y. Yang, D. Liu, A. Liang, Q. Xu, J. Kroder, V. S\"{u}$\beta$, H. Borrmann, C. Shekhar, Z. Wang, C. Xi, W. Wang, W. Schnelle,  S. Wirth, Y.-L. Chen, S. T. B. Goennenwein, and C. Felser, Nat. Phys. \textbf{14}, 1125-1131 (2018).
\bibitem{MoraliScience2019} N. Morali, R. Batabyal, P. K. Nag, E. Liu, Q. Xu, Y. Sun, B. Yan, C. Felser, N. Avraham, and H. Beidenkopf, Science \textbf{365}, 1286-1291 (2019).
\bibitem{LiuScience2019} D. F. Liu, A. J. Liang, E. K. Liu, Q. N. Xu, Y. W. Li, C. Chen, D. Pei, W. J. Shi, S. K. Mo, P. Dudin, T. Kim, C. Cacho, G. Li, Y. Sun, L. X. Yang, Z. K. Liu, S. S. P. Parkin, C. Felser, and Y. L. Chen, Science \textbf{365}, 1282-1285 (2019).
\bibitem{IsakovPRL2006} S. V. Isakov, S. Wessel, R. G. Melko, K. Sengupta, and Y. B. Kim, Phys. Rev. Lett. \textbf{97}, 147202 (2006).
\bibitem{WangPRB2013} W.-S. Wang, Z.-Z. Li, Y.-Y. Xiang, and Q.-H. Wang, Phys. Rev. B \textbf{87}, 115135 (2013).
\bibitem{GuoPRB2009} H.-M. Guo and M. Franz, Phys. Rev. B \textbf{80}, 113102 (2009).
\bibitem{OBrienPRB2010} A. O’Brien, F. Pollmann, and P. Fulde, Phys. Rev. B \textbf{81}, 235115 (2010).
\bibitem{RueggPRB2011} A. R\"{u}egg and G. A. Fiete, Phys. Rev. B \textbf{83}, 165118 (2011).
\bibitem{KoPRB2009} W.-H. Ko, P. A. Lee, and X.-G. Wen, Phys. Rev. B \textbf{79}, 214502 (2009).
\bibitem{KieselPRB2012} M. L. Kiesel and R. Thomale, Phys. Rev. B \textbf{86}, 121105(R) (2012).
\bibitem{KieselPRL2013} M. L. Kiesel, C. Platt, and R. Thomale, Phys. Rev. Lett. \textbf{110}, 126405 (2013).
\bibitem{OrtizPRM2019} B. R. Ortiz, L. C. Gomes, J. R. Morey, M. Winiarski, M. Bordelon, J. S. Mangum, I. W. H. Oswald, J. A. Rodriguez-Rivera, J. R. Neilson, S. D. Wilson, E. Ertekin, T. M. McQueen, and E. S. Toberer, Phys. Rev. Mater. \textbf{3}, 094407 (2019).
\bibitem{OrtizPRL2020} B. R. Ortiz, S. M. L. Teicher, Y. Hu, J. L. Zuo, P. M. Sarte, E. C. Schueller, A. M. Milinda Abeykoon, M. J. Krogstad, S. Rosenkranz, R. Osborn, R. Seshadri, L. Balents, J. He, and S. D. Wilson, Phys. Rev. Lett. \textbf{125}, 247002 (2020).
\bibitem{OrtizPRM2021} B. R. Ortiz, P. M. Sarte, E. M. Kenney, M. J. Graf, S. M. L. Teicher, R. Seshadri, and S. D. Wilson, Phys. Rev. Mater. \textbf{5}, 034801 (2021).
\bibitem{LiAX2021} H. X. Li, T. T. Zhang, T. Yilmaz, Y. Y. Pai, C. Marvinney, A. Said, Q. Yin, C. Gong, Z. Tu, E. Vescovo, R. G. Moore, S. Murakami, H. C. Lei, H. N. Lee, B. Lawrie, and H. Miao, arXiv:2103.09769 (2021).
\bibitem{LiangAX2021} Z. Liang, X. Hou, W. Ma, F. Zhang, P. Wu, Z. Zhang, F. Yu, J.-J. Ying, K. Jiang, L. Shan, Z. Wang, and X.-H. Chen, arXiv:2103.04760 (2021).
\bibitem{LiuAX2021} Z. Liu, N. Zhao, Q. Yin, C. Gong, Z. Tu, M. Li, W. Song, Z. Liu, D. Shen, Y. Huang, K. Liu, H. Lei, and S.-C. Wang, arXiv:2104.01125 (2021).
\bibitem{YangSA2020} S.-Y. Yang, Y. Wang, B. R. Ortiz, D. Liu, J. Gayles, E. Derunova, R. Gonzalez-Hernandez, L. \v{S}mejkal, Y. Chen, S. S. P. Parkin, S. D. Wilson, E. S. Toberer, T. McQueen, and M. N. Ali, Sci. Adv. \textbf{6}, eabb6003 (2020).
\bibitem{NandkishoreNP2012} R. Nandkishore, L. S. Levitov, and A. V. Chubukov, Nat. Phys. \textbf{8}, 158-163 (2012).
\bibitem{TanAX2021} H. Tan, Y. Liu, Z. Wang, and B. Yan, arXiv:2103.06325 (2021).
\bibitem{DuanAX2021} W. Duan, Z. Nie, S. Luo, F. Yu, B. R. Ortiz, L. Yin, H. Su, F. Du, A. Wang, Y. Chen, X. Lu, J. Ying, S. D. Wilson, X. Chen, Y. Song, and H. Yuan, arXiv:2103.11796 (2021).
\bibitem{CZhaoAX2021} C. C. Zhao, L. S. Wang, W. Xia, Q. W. Yin, J. M. Ni, Y. Y. Huang, C. P. Tu, Z. C. Tao, Z. J. Tu, C. S. Gong, H. C. Lei, Y. F. Guo, X. F. Yang, and S. Y. Li, arXiv:2102.08356 (2021).
\bibitem{Xiang AX2021} Y. Xiang, Q. Li, Y. Li, W. Xie, H. Yang, Z. Wang, Y. Yao, and H. H. Wen, arXiv:2104.06909 (2021).
\bibitem{JiangAX2020} Y.-X. Jiang, J.-X. Yin, M. M. Denner, N. Shumiya, B. R. Ortiz, J. He, X. Liu, S. S. Zhang, G. Chang, I. Belopolski, Q. Zhang, M. S. Hossain, T. A. Cochran, D. Multer, M. Litskevich, Z.-J. Cheng, X. P. Yang, Z. Guguchia, G. Xu, Z. Wang, T. Neupert, S. D.Wilson, and M. Z. Hasan, arXiv:2012.15709 (2020).
\bibitem{ChenAX2021} H. Chen, H. Yang, B. Hu, Z. Zhao, J. Yuan, Y. Xing, G. Qian, Z. Huang, G. Li, Y. Ye, Q. Yin, C. Gong, Z. Tu, H. Lei, S. Ma, H. Zhang, S. Ni, H. Tan, C. Shen, X. Dong, B. Yan, Z. Wang, and H.-J. Gao, arXiv:2103.09188 (2021).
\bibitem{HZhaoAX2021} H. Zhao, H. Li, B. R. Ortiz, S. M. L. Teicher, T. Park, M. Ye, Z. Wang, L. Balents, S. D. Wilson, and I. Zeljkovic, arXiv:2103.03118 (2021).
\bibitem{ZhouAX2021} X. Zhou, Y. Li, X. Fan, J. Hao, Y. Dai , Z. Wang , Y. Yao, and H.-H. Wen, arXiv:2104.01015 (2021).
\bibitem{SM} See Supplemental Material for details of the experimental condition and the band-calculation method, the normal-emission measurements, the SOC-induced Dirac-gap opening, the hump structure at 70 meV in the CDW phase, $k_z$-broadening effect, and a full set of ARPES intensities and EDCs for the CDW-gap measurements.
\bibitem{ZWangAX2021} Z. Wang, Y.-X. Jiang, Y. Li, J.-X. Yin, G.-Y. Wang, H.-L. Huang, J. Liu, P. Zhu, N. Shumiya, M. S. Hossain, H. Liu, Y. Shi, J. Duan, X. Li, G. Chang, P. Dai, H. Zheng, J. Jia, M. Z. Hasan, and Y. Yao, Phys. Rev. B \textbf{104}, 075148 (2021).
\bibitem{Blaha2013} P. Blaha, K. Schwarz, G. K. H. Madsen, D. Kvasnicka and J. Luitz, WIEN2K, An Augmented Plane Wave + Local Orbitals Program for Calculating Crystal Properties (Karlheinz Schwarz, Technische Universitat Wien, Austria 2013).
\bibitem{PerdewPRL1996} J. P. Perdew, K. Burke, and M. Ernzerhof, Phys. Rev. Lett. \textbf{77}, 3865 (1996).
\bibitem{PerdewPRL1997} J. P. Perdew, K. Burke, and M. Ernzerhof, Phys. Rev. Lett. \textbf{78}, 1396 (1997).
\bibitem{JZhaoAX2021} J. Zhao, W. Wu, Y. Wang, and S. A. Yang, arXiv:2103.15078 (2021).
\bibitem{YuAX2021} F. H. Yu, T. Wu, Z. Y. Wang, B. Lei, W. Z. Zhuo, J. J. Ying, and X. H. Chen, arXiv:2102.10987 (2021). 
\bibitem{FengAX2021} X. Feng, K. Jiang, Z. Wang, and J. Hu, arXiv:2103.07097 (2021).
\bibitem{WangAX2021} Z. Wang, S. Ma, Y. Zhang, H. Yang, Z. Zhao, Y. Ou, Y. Zhu, S. Ni, Z. Lu, H. Chen, K. Jiang, L. Yu, Y. Zhang, X. Dong, J. Hu, H.-J. Gao, and Z. Zhao, arXiv:2104.05556 (2021).

\end{thebibliography}

\clearpage
\section{Supplemental Material}
\subsection{Experimental condition and band-calculation method}
High-quality single crystals of CsV$_3$Sb$_5$ were synthesized with the self-flux method \cite{OrtizPRM2019, ZWangAX2021}. ARPES measurements were performed using Scienta-Omicron SES2002, DA30, and SES2002 spectrometers at Tohoku University, BL-28A, and BL-2A in Photon Factory (PF), KEK, respectively. We used the He discharge lamp ($h\nu$ = 21.218 eV) at Tohoku University and energy tunable photons of the vacuum ultraviolet (VUV) region ($h\nu$ = 85-125 eV) at BL-28A and the soft-x-ray (SX) region ($h\nu$ = 230-350 eV) at BL-2A in PF. The energy resolution ($\Delta E$) was set to be 7-30 meV at Tohoku University, and 35-100 meV at PF. Specifically, the ARPES data shown in Figs. 2(e)-2(j), 3(b)-3(d), and Supplementary Figs. 8(e) and 9 were obtained at Tohoku University with $\Delta E$ = 7 meV, those in Supplementary Figs. 8(a)-8(d) were obtained at Tohoku University with $\Delta E$ = 30 meV, those in Figs. 1, 2(a)-2(d), 3(a), and Supplementary Figs. 5(a), 5(b), 6, 7, 8(f), and 8(g) were obtained at BL-28A in PF with $\Delta E$ = 35 meV, and those in Figs. 5(c) and 8(h) were obtained at BL-2A in PF with $\Delta E$ = 100 meV. The angular resolution was set to be 0.2$^{\circ}$ at Tohoku University as well as BL-2A in PF and 0.3$^{\circ}$ at BL-28A in PF. Crystals were cleaved in-situ in an ultrahigh vacuum of $<$1$\times$10$^{-10}$ Torr. $E_{\rm F}$ of the sample was referenced to that of a gold film evaporated onto the sample holder. The first-principles band-structure calculations were carried out using the full-potential linearized augmented plane-wave method implemented in the WIEN2K code \cite{Blaha2013} with generalized gradient approximation (GGA) \cite{PerdewPRL1996} and the Perdew-Burke-Ernzerhof (PBE) \cite{PerdewPRL1997} type exchange-correlation potential. Spin-orbit coupling was included self-consistently, and the lattice parameters were directly obtained from experiments \cite{OrtizPRM2019}. k-points mesh for the irreducible BZ is 17$\times$17$\times$10. The Muffin-tin radii (RMT) are 2.50 a.u. for Cs and V and 2.60 a.u. for Sb, respectively. The maximum modulus for the reciprocal vectors Kmax was chosen to satisfy RMT$\times$Kmax = 8.0.

\begin{figure}
\includegraphics[width=3.4in]{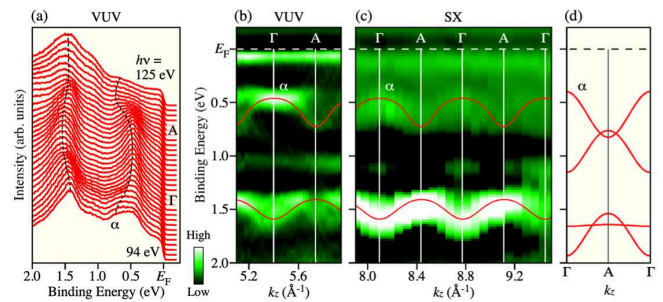}
\vspace{0cm}
\caption{(a) $h\nu$ dependence of EDCs for CsV$_3$Sb$_5$ measured in the normal-emission set up at $T$ = 20 K. (b) Second-derivative intensity of EDCs plotted as a function of $k_z$ and binding energy ($E_{\rm B}$) obtained with VUV photons ($h\nu$ = 94-125 eV). (c) Same as (b) but obtained with the SX photons ($h\nu$ = 230-350 eV). (d) The calculated band structure along the $k_z$ cut obtained from first-principles band-structure calculations.}
\end{figure}

\subsection{Normal-emission measurement}
Figure 5(a) shows energy distribution curves (EDCs) measured with synchrotron radiation in the VUV region at the normal-emission set up, reproduced from Fig. 1(b). One can recognize that the energy location of the $\alpha$-band bottom shows a systematic variation as a function of $h\nu$ and such variation is also visualized in the second-derivative intensity plot in Fig. 5(b). From the periodicity of band dispersion together with the normal-emission ARPES data obtained with SX photons [Fig. 5(c)], we have estimated the inner-potential value to be $V_0$ = 10.0 eV.

\subsection{SOC-induced gap opening on a Dirac band}
Figures 6(a) and 6(b) show the magnified view of the APRES intensity around the Dirac gap and the corresponding EDCs, respectively, in which the gapped nature of Dirac cones is clearly visible.

\begin{figure}
\includegraphics[width=3.4in]{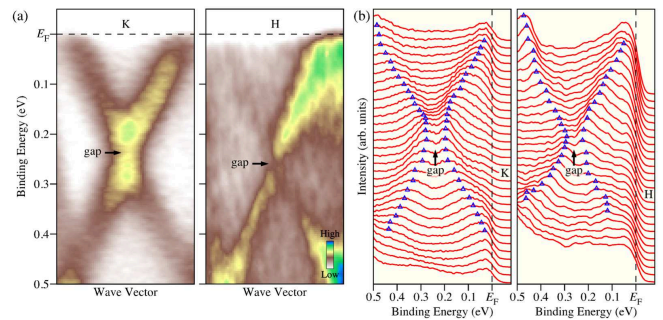}
\vspace{0cm}
\caption{(a) ARPES intensity for the Dirac-cone dispersions at the K (left) and H (right) points. (b) Corresponding EDCs of (a). Blue triangles are a guide for the eyes to trace the peak position of EDCs.}
\end{figure}

\begin{figure}
\includegraphics[width=2.6in]{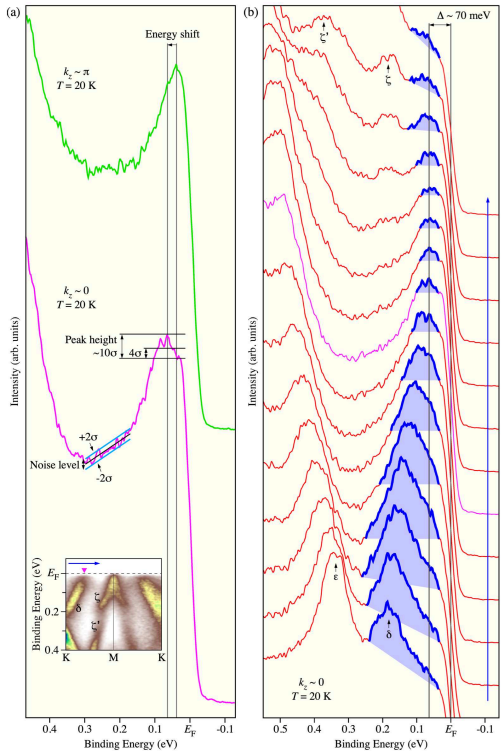}
\vspace{0cm}
\caption{(a) Comparison of EDCs at $T$ = 20 K at the $k_{\rm F}$ point of the $\delta$ band ($k_z$ = 0; magenta) and the $\gamma$ band ($k_z$ = $\pi$; green). Black and light blue lines at the binding energy of 0.2-0.3 eV are the linear fit and those with an offset of $\pm$2$\sigma$, respectively. Inset shows the ARPES intensity along the KMK cut at $T$ = 20 K. (b) EDCs measured along the \textbf{k} cut crossing the $k_{\rm F}$ point of the $\delta$ band [blue arrow shown in the inset to (a)]. The peak derived from the $\delta$ band is highlighted with blue shade.}
\end{figure}

\subsection{Hump structure in the CDW phase}
In Fig. 7(a), we reproduce two EDCs from Fig. 2(c) in the main text; the magenta and green curves were recorded at 20 K at $k_z$ = 0 and $\pi$, respectively. The peak position of the magenta curve ($\sim$70 meV) is shifted toward higher binding energy compared with that of the green curve ($\sim$40 meV), indicative of a CDW-gap opening at $k_z$ = 0 (note that the CDW gap at $k_z$ = $\pi$ is not visible due to the limited resolution of synchrotron-based ARPES). To confirm that the observed peak-energy shift is not due to noise, we have evaluated the noise level of the spectrum. We have estimated the standard deviation ($\sigma$) in the linear fit to a feature-less energy region (black line) and obtained the noise level of $\pm$2$\sigma$ (see light blue lines; note that the linear fit would give the upper limit for the noise level because the inclusion of additional parameters commonly gives a better fit). It is remarked here that the peak height of the magenta curve [roughly 10$\sigma$, see Fig. 7(a)], as defined by the difference in the spectral intensity at 70 meV and 40 meV, is more than twice larger than the noise level (4$\sigma$). This demonstrates that the observed peak-energy shift is not an artifact due to noise, but reflects the CDW-gap opening. To further corroborate the CDW-gap opening, we plot in Fig. 7(b) the EDCs measured at 20 K along a \textbf{k} cut including the $k_{\rm F}$ point [blue arrow in the inset to Fig. 7(a)]. One can see that the $\delta$ band (blue shade) dispersing from the high binding energy side reaches the binding energy of 70 meV at $k_{\rm F}$ (magenta curve) and then almost flattens, in agreement with the CDW-gap opening in the $\delta$ band.

\begin{figure}
\includegraphics[width=3.4in]{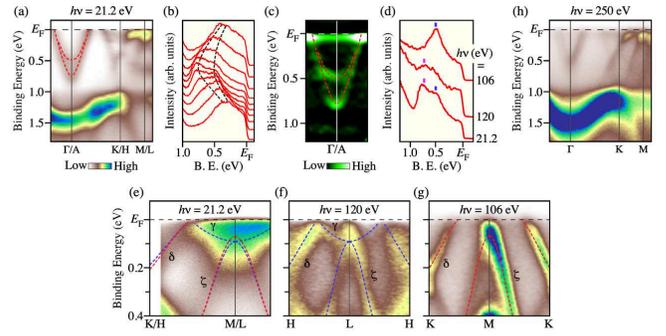}
\vspace{0cm}
\caption{(a) ARPES intensity measured at 25 K along the $\Gamma$KM/AHL line with $h\nu$ = 21.2 eV. (b) Magnified view of the EDCs around the $\alpha$-band bottom at the $\Gamma$/A point. (c) Second-derivative intensity around the $\Gamma$/A point. Dashed curves in (a)-(c) are a guide for the eyes to trace the $\alpha$-band dispersion. (d) Comparison of the normal-emission EDCs measured with $h\nu$ = 106, 120, and 21.2 eV. (e)-(g) ARPES-intensity plots measured along the MK/LH cut obtained with $h\nu$ = 21.2 eV, 120 eV, and 106 eV, respectively. Red and blue dashed curves are a guide for the eyes to trace the band dispersions at $k_z$ = 0 and $\pi$, respectively. (h) ARPES intensity measured along the $\Gamma$KM cut with $h\nu$ = 250 eV.}
\end{figure}

\subsection{$k_z$-broadening effect}
Figure 8(a) shows the ARPES intensity measured along the $\Gamma$KM/AHL cut with the He-I$\alpha$ photons ($h\nu$ = 21.2 eV). One can identify electron-like band ($\alpha$ band) centered at the $\Gamma$/A point. A closer look at the EDCs around the $\Gamma$/A point reveals that the $\alpha$ band consists of two branches each bottomed at $E_{\rm B}$ of $\sim$0.5 and $\sim$0.75 eV [see Fig. 8(b); also see the second-derivative intensity around the $\Gamma$/A point in Figs. 8(c)]. These branches are attributed to the $\alpha$ band at the $k_z$ = 0 ($\Gamma$KM) plane and at the $k_z$ = $\pi$ (AHL) plane, respectively, as supported by the observation of the $\alpha$-band bottom at $E_{\rm B}$ $\sim$ 0.5 eV at $k_z$ = 0 and 0.75 eV at $k_z$ = $\pi$ [Fig. 8(d); also see Figs. 1(c) and 1(d)] which agrees well with the energy position of the two peaks at $h\nu$ = 21.2 eV. Considering short photoelectron escape depth (5-10 \AA) of 21.2-eV photons and a resultant large $k_z$ broadening effect, the ARPES data reflect the photoelectron signal averaged over a wide $k_z$ area in the bulk BZ so that the band structures at $k_z$ = 0 and $\pi$ planes which have a large contribution to the density of states are simultaneously observed. Such a $k_z$ broadening effect is also seen around the Brillouin-zone corner. Namely, the band dispersions observed along the MK/LH cut with 21.2-eV photons can be explained by a simultaneous observation of the bulk bands at $k_z$ = 0 and $\pi$ [see Figs. 8(e)-8(g)]. It is noted that overall band dispersions in a wide energy range and their ARPES-intensity distributions obtained with He-I$\alpha$ photons [Fig. 8(a)] also agree with those for the bulk bands observed with SX photons [Fig. 8(h)] except for the appearance of the second $\alpha$-band branch, supporting the bulk origin of the observed bands. Here, we briefly comment on the amount of the $k_z$ dispersion. Since the $\alpha$-band bottom is located at $\sim$0.5 eV and $\sim$0.75 eV at $k_z$ = 0 and $\pi$, respectively, the total $k_z$ dispersion is $\sim$0.25 eV in the experiment. This is smaller than the calculated value of $\sim$0.4 eV, suggesting that the $k_z$ dispersion is overestimated in the calculation.

\begin{figure}
\includegraphics[width=3.4in]{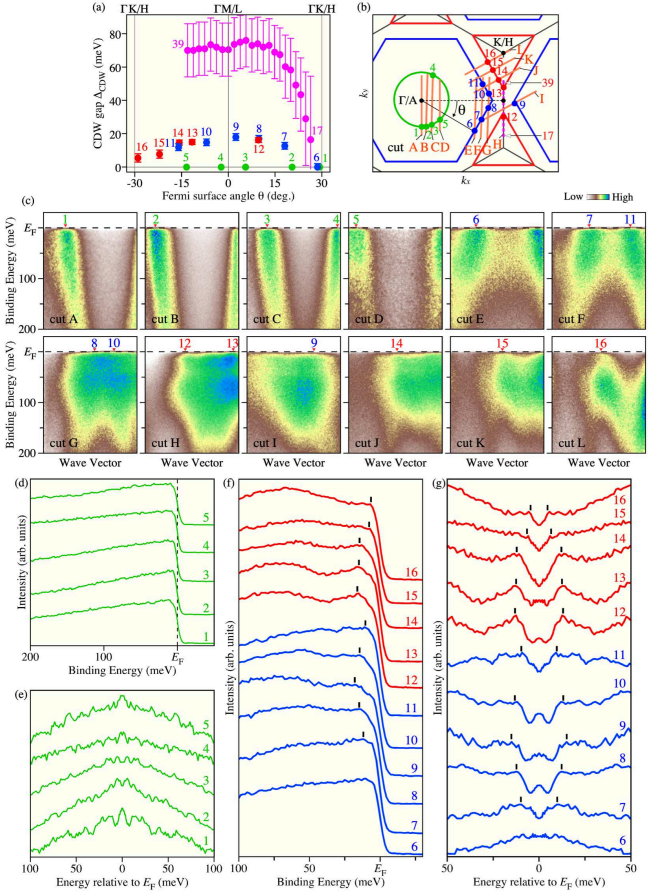}
\vspace{0cm}
\caption{(a) Plots of the CDW-gap magnitude $\Delta_{\rm CDW}$ as a function of $\theta$. Red, blue/light blue, and green circles represent $\Delta_{\rm CDW}$ for the saddle-point band, Dirac band, and the $\alpha$ band. (b) Schematic of the Fermi surface together with the \textbf{k} locations where $\Delta_{\rm CDW}$ was estimated (red, blue, and green circles). Definition of $\theta$ is also shown. (c) ARPES intensity at 10 K along cuts A-L in (b). (d), (e) EDCs and symmetrized EDCs at 10 K, respectively, at points 1-5. (f), (g) Same as (d) and (e), respectively, but for points 6-16.}
\end{figure}

\subsection{ }

\subsection{Full data set for the CDW-gap measurements}
Figure 9(a) shows the gap plot reproduced from Fig. 3(e) but additionally includes the numbering of the k points at which the gap size has been estimated [see Fig. 9(b)]. Figure 9(c) shows the ARPES intensity plots at $T$ = 10 K along various \textbf{k} cuts with different Fermi surface angles ($\theta$) [orange lines A-L in Fig. 9(b)]. One can see a general trend that the band dispersion around the M point ($\theta$ = 0$^{\circ}$) is flattened by an opening of a large CDW gap (e.g, cuts G, H, and J), while a relatively steep dispersion is retained around the $\Gamma$K cut ($\theta$ = 30$^{\circ}$; e.g., E, F, and L) due to a small CDW gap. Figures 9(d)-9(g) display EDCs measured at various $k_{\rm F}$ points indicated by red triangles in Fig. 9(c), for the $\alpha$ band [Fig. 9(d)] and the Dirac band [Fig. 9(f)], as well as the corresponding symmetrized EDCs [Figs. 9(e) and 9(g) for the $\alpha$ and Dirac band, respectively]. For the $\alpha$ band, one can recognize the absence of a clear gap opening irrespective of $\theta$. For the Dirac band, blue curves in Figs. 9(e) and 9(f) were obtained on the large hexagonal Fermi surface centered at the $\Gamma$/A point, whereas red ones were obtained on the triangular Fermi surface centered at the K/H point. The gap size estimated on these Fermi surfaces is almost the same in magnitude and shows a similar anisotropy; it shows the maximum of 16-18 meV along the $\Gamma$M/AL line ($\theta$ = 0$^{\circ}$), while it shows the minimum of 0-5 meV along the $\Gamma$K/AH line ($\theta$ = 30$^{\circ}$). EDCs for the saddle-point band are displayed in Fig. 2(g) of the main text.

\end{document}